\documentclass[twocolumn,superscriptaddress,amsmath,amssymb,aps,prl,10pt]{revtex4-1}

\usepackage{graphicx}
\usepackage{hyperref}

\begin{document}

\title{Chirality dependence of the absorption cross-section of carbon nanotubes.}

\author{Fabien Vialla}
\affiliation{Laboratoire Pierre Aigrain, \'Ecole Normale Sup\'erieure, UPMC, Universit\'e Paris Diderot, CNRS, 75005 Paris, France}
\author{Cyrielle Roquelet}
\altaffiliation{Now at Columbia University, NY, USA}
\affiliation{Laboratoire Aim\'e Cotton, \'Ecole Normale Sup\'erieure de Cachan, Universit\'e Paris Sud, CNRS, 91405 Orsay, France}
\author{Benjamin Langlois}
\affiliation{Laboratoire Pierre Aigrain, \'Ecole Normale Sup\'erieure, UPMC, Universit\'e Paris Diderot, CNRS, 75005 Paris, France}
\author{G\'eraud Delport}
\affiliation{Laboratoire Aim\'e Cotton, \'Ecole Normale Sup\'erieure de Cachan, Universit\'e Paris Sud, CNRS, 91405 Orsay, France}
\author{Silvia Morim Santos}
\affiliation{Laboratoire Aim\'e Cotton, \'Ecole Normale Sup\'erieure de Cachan, Universit\'e Paris Sud, CNRS, 91405 Orsay, France}
\author{Emmanuelle Deleporte}
\affiliation{Laboratoire Aim\'e Cotton, \'Ecole Normale Sup\'erieure de Cachan, Universit\'e Paris Sud, CNRS, 91405 Orsay, France}
\author{Philippe Roussignol}
\affiliation{Laboratoire Pierre Aigrain, \'Ecole Normale Sup\'erieure, UPMC, Universit\'e Paris Diderot, CNRS, 75005 Paris, France}
\author{Claude Delalande}
\affiliation{Laboratoire Pierre Aigrain, \'Ecole Normale Sup\'erieure, UPMC, Universit\'e Paris Diderot, CNRS, 75005 Paris, France}
\author{Christophe Voisin}
\email{christophe.voisin@lpa.ens.fr}
\affiliation{Laboratoire Pierre Aigrain, \'Ecole Normale Sup\'erieure, UPMC, Universit\'e Paris Diderot, CNRS, 75005 Paris, France}
\author{Jean-S\'ebastien Lauret}
\affiliation{Laboratoire Aim\'e Cotton, \'Ecole Normale Sup\'erieure de Cachan, Universit\'e Paris Sud, CNRS, 91405 Orsay, France}
\date{\today}

\begin{abstract}
The variation of the optical absorption of carbon nanotubes with their geometry has been a long standing question at the heart of both metrological and applicative issues, in particular because optical spectroscopy is one of the primary tools for the assessment of the chiral species abundance of samples. Here, we tackle the chirality dependence of the optical absorption with an original method involving ultra-efficient energy transfer in porphyrin/nanotube compounds that allows uniform photo-excitation of all chiral species. We measure the absolute absorption cross-section of a wide range of semiconducting nanotubes at their $S_{22}$ transition and show that it varies by up to a factor of 2.2 with the chiral angle, with type I nanotubes showing a larger absorption. In contrast, the luminescence quantum yield remains almost constant.

\end{abstract}

\maketitle


The versatility of the physical properties of Single-Wall carbon Nanotubes (SWNTs) with respect to their geometry (the so-called $(n,m)$ chiral species) is very attractive for applications \cite{Avouris2008, Lauret2004b, Kostarelos2009, Liu2009}, but on the other hand, the uncontrolled mixtures of species produced by regular synthesis methods blur out their specific properties. 
Post-growth sorting methods now allow to enrich samples in some specific species \cite{Arnold2006}, but they also miss a tool for the quantitative assessment of their outcome. 
Optical techniques such as absorption, photoluminescence (PL) or resonant Raman spectroscopies are the primary tools to this end. However, these techniques can neither give a quantitative estimate of the species concentration nor their relative abundance to-date because they miss the knowledge of the ($n,m$)-dependence of the optical cross-section at the nanotubes' resonances ($S_{11}$ and $S_{22}$). Although several studies pointed that the optical properties of carbon nanotubes depend on their chiral angle, they all actually dealt with a combination of physical parameters (such as absorption cross-section, Raman scattering cross-section or PL quantum efficiency). As a result, the literature gives quite contradictory or inconclusive results, some of them pointing to a larger abundance of near armchair nanotubes (interpreted as energetically favored in the growth process) whereas other studies concluded for a larger optical cross-section for large chiral angles \cite{Bachilo2002,Miyauchi2004,Okazaki2006,
Jorio2006,Luo2006}.

\begin{figure}[h]
	\centering
		\includegraphics[width=0.47\textwidth]{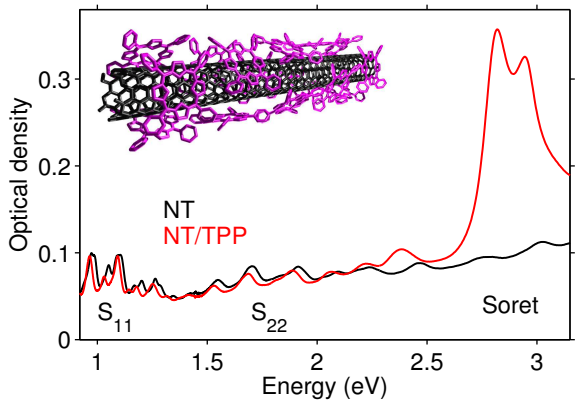}
	\caption{Optical absorption spectra of HiPPCO SWNTs (black) and SWNT/TPP compounds (red) in micellar solutions. Inset : Schematic view of the non covalent SWNT/TPP compound.}
	\label{fig:Abs_Hipco-TPP}
\end{figure}

Here, we propose an original method for assessing the chirality dependence of the absorption cross-section of semiconducting carbon nanotubes, by means of non-covalent functionalization with tetraphenyl porphyrin (TPP) molecules (Inset of Figure~\ref{fig:Abs_Hipco-TPP}). This functionalization gives rise to an extremely efficient energy transfer \cite{Roquelet2010APL} that allows to excite uniformly the whole set of carbon nanotubes regardless of their chirality. By comparison with the PL signal obtained in the regular excitation scheme (on the intrinsic $S_{22}$ transition of the SWNTs) of the same sample, we can single out the contribution of the absorption cross-section in the chiral dependence of the PL intensity. We show that the main variation of this absorption cross-section comes from the chiral angle $\theta$ \footnote{$\cos \theta = \frac{
2n+m}{2\sqrt{n^2+m^2+nm}}$, see \textit{e.g.} \cite{Reich2004}} and fits well to the inverse of the geometrical parameter $q \cos (3\theta)$, where $q = n-m$ (mod 3) stands for the family type: $q=+1$ (resp. $q=-1$) for the so-called type II (resp. type I) nanotubes. In contrast, we show that the PL quantum yield hardly depends on the chiral species. This opens the way to the quantitative analysis of the chiral species content of samples by means of optical tools. 

The $(n,m)$ dependence of the optical cross-section of carbon nanotubes has been investigated theoretically by several teams. Computations by Reich~\textit{et al.}~\cite{Reich2005}, Oyama~\textit{et al.}~\cite{Oyama2006} and Malic \textit{et al.} \cite{Malic2006} suggest with different physical arguments that the $S_{22}$ absorption of the chiral species with small chiral angles and $q=+1$ is intrinsically weaker.
From an experimental point of view, several studies combining PL and Raman spectroscopies \cite{Jorio2006} or PL and TEM \cite{Okazaki2006} concluded for a sizable $(n,m)$ dependence of the optical signals. However, the specific $(n,m)$ dependence of the absorption cross-section $\sigma$ could not be singled out. 
As a workaround, Tsyboulski \textit{et al.} \cite{Tsyboulski2007} proposed to use empirical factors to estimate the abundance of each chirality based on the so-called action cross-section $A$ that combines the absorption cross-section at the $S_{22}$ transition $(\sigma_{S_{22}})$ and the photoluminescence quantum yield $(\phi_{PL})$ of the nanotubes. 
In total, a thorough experimental investigation of the chiral angle dependence of the absorption of SWNTs is still lacking despite the important metrological and applicative issues at stake.



Non purified HiPCO and CoMoCat nanotubes were functionalized with free-base tetraphenyl porphyrin (TPP) in aqueous solution by means of the micelle swelling method (see Ref.~\cite{Roquelet2010ChemPhysChem} for details). 
The optical absorption spectrum of the HiPCO nanotube/porphyrin compounds is shown in Figure~\ref{fig:Abs_Hipco-TPP}.

The resonance at 2.82~eV corresponds to the so-called Soret band of the TPP molecules stacked on the nanotube walls. The shoulder at 2.95~eV is the contribution of residual free porphyrins. The absorption bands in the 0.9-1.35~eV range correspond to the $S_{11}$ transitions of the various chiral species of nanotubes.

\begin{figure}[h!]
	\centering
		\includegraphics[width=0.50\textwidth]{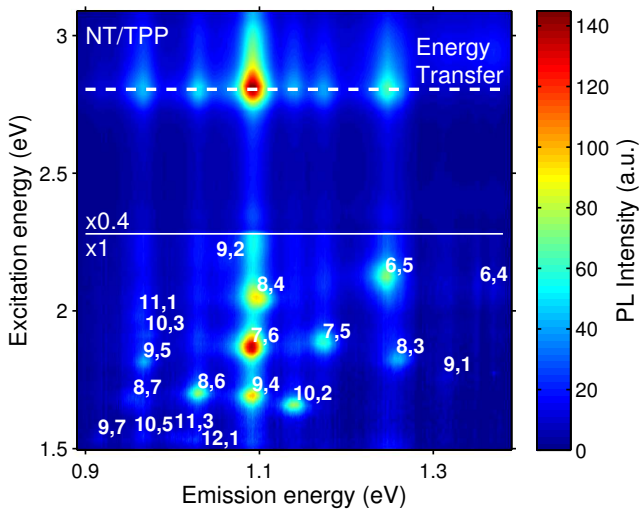}
	\caption{PL map of the HiPCO SWNT/TPP compounds suspension. The dashed white line at 2.82~eV is a guide to the eye showing the energy transfer resonance upon excitation of the TPP molecules. The spectra are normalized to a constant incoming photon flux. The upper part intensities are reduced by a factor 0.4 for the sake of clarity.}
	\label{fig:Map-2D}
\end{figure}

PL maps were recorded using an InGaAs detector and using the output of a monochromator illuminated by a UV-vis Xe lamp for the excitation (5~nm excitation steps). The PL intensity is normalized by the incoming photon flux at each excitation wavelength. The lower part of the PL map (Figure~\ref{fig:Map-2D}) displays several bright spots that correspond to the emission of carbon nanotubes at their $S_{11}$ transition upon excitation on their $S_{22}$ transition. Each spot can be assigned to a specific chiral species following the procedure proposed by Bachilo \textit{et al.}~\cite{Bachilo2002}. An additional set of resonances can be seen at the same emission energies for an excitation at 2.82~eV. This energy corresponds to the absorption of the porphyrin molecules stacked on the nanotube. We assign these spots to the resonant excitation of the Soret transition of porphyrin followed by energy transfer to the nanotube and by the regular $S_{11}$ emission of the nanotubes 
\cite{Roquelet2010APL,magadur2008,casey2008,Sprafke2011}. In other words, the emission of the nanotubes is enhanced when the excitation is tuned in resonance with the porphyrin molecules. As can be seen qualitatively in the figure, this resonance appears for all chiral species. Therefore, this energy transfer resonance provides a new handle to achieve uniform photo-excitation of the whole set of chiral species.\\

This PL map allows us to compare the optical properties of the different chiral species of nanotubes and infer their intrinsic absorption cross-sections. 
Let us define, for each $(n,m)$ species, the ratio $R$ between the PL intensities recorded for an excitation on the Soret resonance ($I^{NT}_{Soret}$) or for an excitation on the intrinsic $S_{22}$ resonance ($I^{NT}_{S_{22}}$) \footnote{Note that the PL due to the intrinsic absorption of SWNTs at 2.82~eV represents about 15\% of the total signal and was subtracted from $I^{NT}_{Soret}$ for the calculation of $R$. At the same time, we took into account the contribution of the nearby $S_{33}$ transition for the (9,7) and (8,7) species.}. Provided that all spectra are normalized to the incoming photon flux, $R$ reads :

\begin{equation}
  R= \frac{I^{NT}_{Soret}}{I^{NT}_{S_{22}}}=\frac{\sigma_{TPP}.N.\eta_{T}.C_{n,m}.\phi_{PL}}{\sigma_{S_{22}}.C_{n,m}.\phi_{PL}} = \frac{\sigma_{TPP}.N.\eta_{T}}{\sigma_{S_{22}}} \label{eq:R}
\end{equation}
$I^{NT}_{Soret}$ is proportional to $\sigma_{TPP}$ (absorption cross-section of the TPP molecule), $N$ (number of molecules stacked on a nanotube per unit length), $\eta_{T}$ (energy transfer quantum yield), $C_{n,m}$ (species concentration) and $\phi_{PL}$ (PL quantum yield of nanotubes). $I^{NT}_{S_{22}}$ is proportional to $\sigma_{S_{22}}$ (absorption cross-section of the nanotube at the $S_{22}$ transition per unit length), $C_{n,m}$ and $\phi_{PL}$~\cite{Roquelet2010APL}.

The important point here, is that $R$ is the ratio of two PL intensities measured on the same transition ($S_{11}$) for the same chiral species and for the same sample. Thus, this ratio allows to eliminate the contribution of the unknown PL quantum yield  $\phi_{PL}$ and the contribution of the unknown species concentration $C_{n,m}$. This point is crucial since these two parameters are very difficult to measure, which is the main reason that has hampered the determination of $\sigma_{S_{22}}$ in previous studies.

Finally, we have shown recently that both the direct and transfer excitation mechanisms share the same polarization diagram due to the reshaping of the electric field in the close vicinity of the nanotube \cite{Roquelet2012}. As a consequence, polarization dependences are also eliminated in $R$ provided that polarized cross-sections are used in the calculation. Namely we obtain $\sigma_{S_{22}}^{//} \simeq 3 <\sigma_{S_{22}}>_{or}$ by using $\sigma_{TPP}^{//} \simeq \frac{3}{2} <\sigma_{TPP}>_{or}$ in (\ref{eq:R}), where $<>_{or}$ stands for the average over random orientations (see SI).

$R$ can bring an original insight into the $(n,m)$ dependence of $\sigma_{S_{22}}$ provided that both the coverage $N$ and the transfer yield $\eta_T$ do not depend on the chiral species (see Eq.~\ref{eq:R}). 
To assess the coverage $N$, we performed in a previous study a systematic analysis of the functionalization degree as a function of the amount of TPP molecules (\cite{Roquelet2010ChemPhysChem} and SI). We found that all the spectroscopic signatures of functionalization reach a plateau above a critical TPP concentration while those of free TPP grow linearly.
This is interpreted as the completion of a full single layer of TPP molecules on the nanotube. Molecular simulations show that the TPP/SWNT distance is 0.32~nm and that the average distance between TPP molecules is $L_{TPP} \simeq 1.6$~nm \cite{Correa2012}. Therefore, only 3 TPP molecules can fit on the circumference of a nanotube for any of the species investigated in this study. Thus, we assume $N=3/L_{TPP}$, with no $(n,m)$ dependence \footnote{W. Orellana, private communication. Note that a coverage proportional to the diameter could also be considered. This would yield variations of $\pm$20\% in the reported values which is of the order of the error bars and would not change the conclusions discussed in the following.}.

The transfer quantum yield $\eta_T$ was assessed in a former study and turned out to be of the order of 99.99\% on average in a sample enriched in the (6,5) species \cite{Roquelet2010APL}. One of methods used for this estimate relies on the average quenching of the $Q$ bands luminescence of TPP molecules stacked on the nanotubes which is greater than $10^3$. We observe the same quenching for unsorted nanotubes. Clearly, this would not be possible if the transfer yield was to be significantly lower for some chiral species.
Therefore, we can safely state that the energy transfer occurs with an almost 100\% efficiency for all chiral species observed in this sample. Note also that the time-resolved measurements reported in Ref.~\cite{Garrot2011} support the same conclusion and give similar results for unsorted nanotubes.

We end up with the important result that $R$ is simply proportional to $1/\sigma_{S_{22}}$, the proportionality coefficient being the same for all chiral species. $R$ is thus a direct image of the relative variations of $\sigma_{S_{22}}$ with the chiral species.

Practically, $R$ is estimated from a global fitting of the lines of the PL map (Figure~\ref{fig:Map-2D}, see SI for details).  We were able to measure $R$ for 13 chiral species spanning a wide range of chiral angles and the 0.68 - 1.1~nm diameter range.

We find that $R$, and hence $\sigma_{S_{22}}$, shows no clear dependence with respect to the nanotube diameter (see SI).

\begin{figure}[h!]
	\centering
		\includegraphics[width=0.50\textwidth]{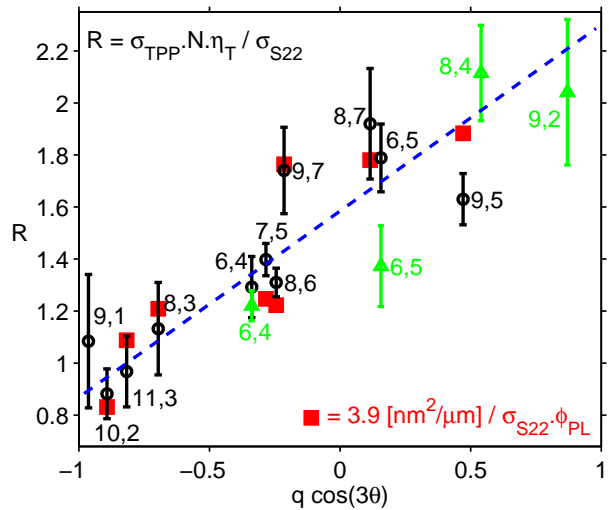}
	\caption{Black circles: $R$ ratio between the PL intensity excited on the Soret resonance and the PL intensity excited on the intrinsic $S_{22}$ resonance in the HiPCO based SWNT/TPP compounds as a function of $q \cos(3\theta)$. Green triangles : same ratio $R$ evaluated for a sample made from CoMoCat material. 
Blue line: linear fit of $R$. Red squares: Inverse of the action cross-section $A$ of pristine nanotubes (from 
Ref~\cite{Tsyboulski2007}) scaled with a simple proportionality factor.}
	\label{fig:efficacite-chiralite-R_AC}
\end{figure}

In contrast, Figure~\ref{fig:efficacite-chiralite-R_AC} shows a linear relationship between $R$ and the geometrical parameter $q\cos(3\theta)$. $R$ and hence $\sigma_{S_{22}}$ strongly depend on the chiral angle regardless of the diameter and vary by up to a factor 2.2 for zigzag nanotubes of opposite families. Generally speaking, type I SWNTs show a larger absorption than their type II counterparts. 

$R$ can be compared for each chiral species to the PL action cross-section $A$ reported in the literature \cite{Tsyboulski2007}. This quantity is the product of the absorption cross section $\sigma_{S_{22}}$ with the PL quantum yield $\phi_{PL}$. 
The inverse of the action cross section experimentally evaluated for individual pristine nanotubes by Tsyboulski \textit{et al.} is reported in Figure \ref{fig:efficacite-chiralite-R_AC}, simply scaled by an arbitrary factor. Obviously, $R$ and $1/A$ share the same variations with $q \cos (3 \theta)$. This excellent agreement between the two sets of data is particularly remarkable for they stem from different types of samples and were obtained with different setups and methods.
This strongly supports our conclusion that $R$ reflects intrinsic properties of the nanotubes, independently of the excitation scheme through the porphyrins. 

 This new insight into the chiral dependence of the absorption cross-section is extremely valuable for assessing the relative abundance of chiral species in a sample and can be used to revisit the previous chiral species abundance assessment deduced from absorption or PL measurements \cite{Bachilo2002}. The absorption corrected data lead to much more symmetric chiral angle distributions, with no preferential type (I or II) of nanotubes (see SI). Retrospectively, the limited number of type II nanotubes observed in this study may be understood as a consequence of their lower absorption and thus reduced PL signal rather than weaker abundance. We note that even after correction near zigzag nanotubes are found to be less abundant raising questions about the underlying growth mechanism.

\begin{figure}[h]
	\includegraphics[width=0.50\textwidth]{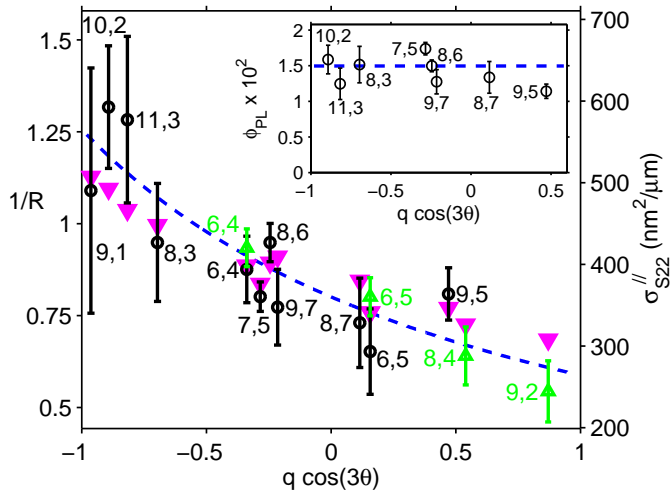}
	\caption{Inverse of the $R$ ratio, which is directly proportional to the absorption cross-section, as a function of $q \cos(3\theta)$ for HiPCO (black dots) and CoMoCat nanotubes (green triangles). Purple down triangles~: Theoretically calculated absorption probability (arb. units, from Ref.~\cite{Oyama2006}). Blue line~: fit to the data leading to the empirical formula~(\ref{eq:emp-law}). Inset~: PL quantum yield as a function of $q \cos(3\theta)$ deduced from the action cross-section measured in Ref.~\cite{Tsyboulski2007} and from the absorption cross-section presented in the main figure.}  
	\label{fig:comp_exp-theory}
\end{figure}

The $(n,m)$ dependence of the absorption of carbon nanotubes can be compared to theoretical models available in the literature. Especially, Oyama \textit{et al.} computed explicitly the absorption cross-section on the $S_{22}$ transition for all the chiral species observed in this study \cite{Oyama2006}. We report their data, simply scaled by an arbitrary factor, in Figure~\ref{fig:comp_exp-theory} together with the experimental absorption cross-sections evaluated in our study. 
The general trend is well accounted for by the calculations, with the largest absorption for $q=-1$, near zigzag nanotubes. In this model, this effect is a consequence of the trigonal warping of the band structure of graphene that leads to a chiral dependent matrix element for optical transitions. However, the calculations give an underestimated variation of the absorption cross section with the chiral angle. This may find its origin in many body effects or $\sigma-\pi$ hybridization effects not included in the model, which can lead to additional $(n,m)$ dependences.
%

Finally, we can extract absolute estimates for the absorption cross-sections, by exploiting the knowledge of the absorption of the TPP molecule used as absorbing unit in the transfer process. Assuming the expressions for $N$ and $\eta_{T}$ discussed previously and an absorption cross-section $\sigma_{TPP}^{//}=2.4 \times 10^{-15}$~cm$^2$ (see SI), we deduce $\sigma_{S_{22}}^{//}$ for the whole set of species (Figure~\ref{fig:comp_exp-theory} right scale and Table in SI). 
In particular, we can compare our estimate for the very few species for which absorption measurements have been reported in the literature, \textit{e.g.} for the $(6,5)$ species \cite{Berciaud2008,Joh2011,Islam2004,Schneck2011,Oudjedi2013}. Using completely different approaches, these studies yielded $\sigma_{S_{22}}^{//}$ ranging from 3 to 300~nm$^2$/$\mu$m, as compared to our own estimate of $\sigma_{S_{22}}(6,5) \simeq 330 \pm 60$~nm$^2$/$\mu$m.

In addition, we can deduce the PL quantum yield $\phi_{PL}$ of each $(n,m)$ species from the scaling factor between $\sigma_{S_{22}}$ and the action cross-section of Ref.~\cite{Tsyboulski2007} (Figure~\ref{fig:efficacite-chiralite-R_AC}). 
We deduce from our absolute estimates of $\sigma_{S_{22}}$ that $\phi_{PL}$ is of the order of 1.4\% in the sample of Ref~\cite{Tsyboulski2007} for all chiral species. The spread of $\phi_{PL}$ is of the order of $\pm$~10\% around this mean value (see Inset of Figure~\ref{fig:efficacite-chiralite-R_AC}). This tends to show that the chiral dependence of the PL quantum yield is negligible in agreement with theoretical predictions \cite{Oyama2006,Malic2006} and more generally that most of the $(n,m)$ dependence in the nanotubes PL signal comes from the $S_{22}$ resonant absorption. Assuming that the non radiative relaxation processes do not strongly depend on the chiral species, we can further infer that the variation of $\sigma_{S_{11}}$ with the chiral species must be much smaller than that of $\sigma_{S_{22}}$.

For practical purposes, we deduce from the linear fit to the data in Figure~\ref{fig:efficacite-chiralite-R_AC}, a simple empirical expression for the absorption cross-section $\sigma_{S_{22}}$ of a $(n,m)$ nanotube :
 \begin{equation}
  \sigma_{S_{22}}^{//} = \frac{10^3}{ q \cos (3 \theta) + 2.8}
 \label{eq:emp-law}
 \end{equation}
where $\sigma$ is in nm$^2/\mathrm{\mu}$m. 
This empirical formula provides an estimate with an uncertainty of about $\pm20$\% and remains restricted to the case of small diameter ($0.7\mathrm{nm}< d_t< 1.1\mathrm{nm}$) semi-conducting nanotubes. A possible generalization would obviously require an additional diameter dependence to be introduced in the empirical formula. This diameter dependence could not be included properly in this study since the diameter variations ($\pm$~20\% around the mean value) are of the order of the experimental uncertainties.

In conclusion, we proposed an original method to address a key pending issue in the physics of carbon nanotubes : the chirality dependence of the absorption cross-section. We measured the absolute variations of this absorption cross-section and proposed an empirical law. In particular, we showed that the type I semi-conducting species show a significantly larger absorption than the type II species, whereas their PL quantum yield are almost identical. This study opens the way to quantitative analysis of the chiral species content of samples based on optical measurements. In addition, the tools developed in this study bring a new attractive feature : a uniform photo-excitation of the whole set of chiral species in a sample. This opens an avenue to the investigation of other intrinsic optical properties of SWNTs and to new 
approaches to the use of SWNTs in labeling applications.

\subsection*{Acknowledgement}
FV and CR participated equally to this work. This work was supported by the
GDR-I GNT, the grant \textit{"C'Nano IdF TENAPO"} and the ANR grant "TRANCHANT". CV is a member of `` Institut Universitaire de France''.\\

\bibliography{biblio-chiralite-sigmaS22}

\part{Supplementary Information}

\section*{time-resolved measurements}

\begin{figure}[h!]
 	\centering
		\includegraphics[width=0.50\textwidth]{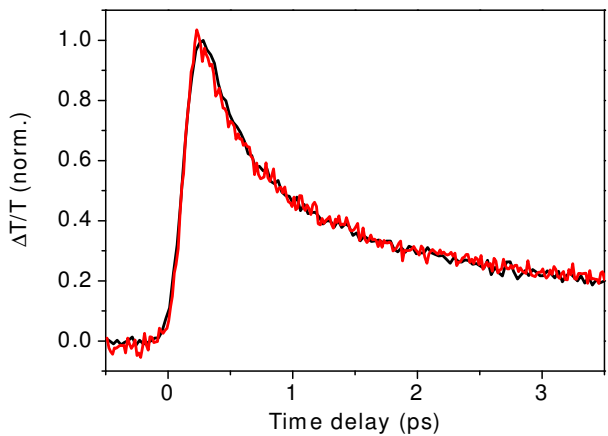}
 	\caption{Normalized transient change of transmission on the $S_{11}$ transition of $(6,5)$ carbon nanotubes subsequent to an optical excitation on the $S_{22}$ transition (red line) or on the excitation energy transfer resonance (Soret band, black line).}
 	\label{fig:ppesde}
 \end{figure}

To compare the two excitation schemes (direct $S_{22}$ excitation or energy transfer from TPP) discussed in this paper, we performed additional time-resolved measurements on SWNT/TPP compounds solutions. We used a two-color femtosecond pump-probe setup based on an optical parametric amplifier. Time-resolved transient absorption measurements are shown in 
Figure~\ref{fig:ppesde} for a probe wavelength in resonance with the $S_{11}$ transition of the (6,5) species for a pump either in resonance with the intrinsic $S_{22}$ transition or with the Soret transition of the TPP (energy transfer resonance). The normalized traces are very similar both for the rise and the decay dynamics. This indicates that the population buildup and the subsequent decay of the $S_{11}$ excitonic level are identical in both excitation schemes. One can therefore rule out different internal conversion dynamics and yields in the two excitation schemes.

\section*{Assessment of the coverage}
\begin{figure}[h!]
 	\centering
		\includegraphics[width=0.50\textwidth]{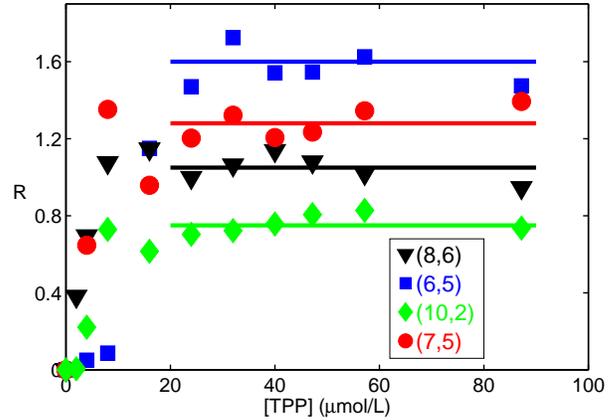}
 	\caption{Transfer ratio $R$ measured for several chiral species as a function of the TPP concentration, showing a saturation above a threshold concentration. The horizontal lines are a guide to the eyes.}
 	\label{fig:RvsTPPconc}
 \end{figure}

The assumption of a full coverage of the nanotubes with a single layer of TPP is supported by two experimental observations. Upon increasing the TPP concentration in the functionalization process, we observe a saturation of the amplitude of the shifted Soret peak (2.82~eV) of the bound TPP, whereas the Soret peak of the free TPP (2.95~eV) keeps increasing (See Fig.3 in \cite{Roquelet2010ChemPhysChem}). This shows that the first layer of TPP is completed. The question of whether additional layers of TPP are built (without giving rise a contribution to the shifted peak) can be addressed by monitoring the transfer ratio $R$ upon increasing the TPP concentration (Fig.~\ref{fig:RvsTPPconc}). Clearly, the energy transfer is not increased for larger TPP concentrations once the first layer is completed. We deduce that even if a second layer is built, it does not participate to the energy transfer, which is a sufficient condition for the $\sigma_{S_{22}}$ assessment scheme developed in this work. In addition, 
molecular simulations show that the binding energy of a TPP molecule on a nanotube ($\simeq1.3$~eV) is much larger than the thermal energy, regardless of the SWNT diameter or chiral angle \cite{Correa2012, Correa2013} supporting the assumption of a full coverage (total reaction). We note however that our method may lead to a small overestimate of $\sigma_{S_{22}}^{//}$ if the coverage compactness is not perfect.

 \section*{Assessment of the TPP absorption cross-section}
 
\begin{figure}[h!]
 	\centering
		\includegraphics[width=0.50\textwidth]{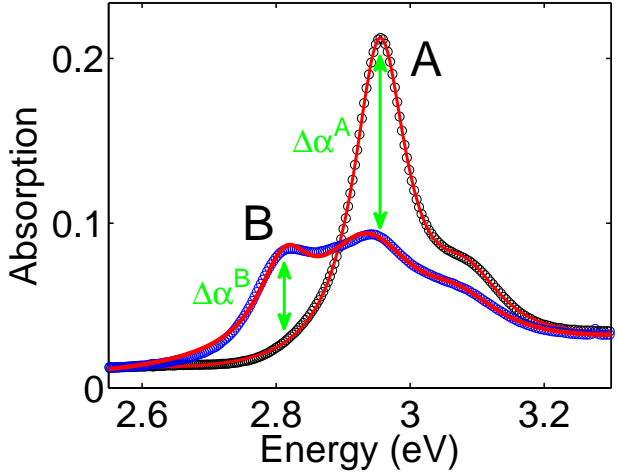}
 	\caption{Absorption spectrum (open dots) and fits (red lines) of the Soret band before the functionalization process (black dots) and after completion of the functionalization (blue dots). The additional component at 2.82~eV corresponds to the bound TPP.}
 	\label{fig:absTPP}
 \end{figure}

 In order to estimate the absorption cross-section of the TPP molecules bound to the nanotubes, we used the following procedure. The micellar suspensions of TPP and SWNT were mixed together and the absorbance was measured right after. At this stage the product consists only in free TPP and NT. The functionalization is then initiated by applying ultrasonic agitation for several hours. After the equilibrium is reached, the absorbance is measured again, showing a split Soret band at 2.95~eV (peak A, free TPP) and 2.82~eV (peak B, bound TPP). The two absorbance spectra are shown in Fig.~\ref{fig:absTPP}.
 In the first stage, the concentration of free TPP is well known and allows us to deduce its average absorption cross-section $<\sigma_{TPP}>_{or}= \frac{2}{3} \sigma_{TPP}^{//} = \frac{\epsilon_{TPP} \ln 10}{10^3 \mathcal{N}_a} =1.6\times 10^{-15}$~cm$^2$, in good agreement with the literature \cite{Kim1972}. In fact, the TPP molecule -that is roughly planar- shows no absorption for an electric field perpendicular to its plane. Therefore, the orientation averaged absorption cross-section of free TPP is 2/3 of the in-plane absorption cross-section $\sigma_{TPP}^{//}$.

 We then make use of a comparison between the two spectra to deduce the absorption of the TPP bound to the nanotube. The decrease of absorbance of the peak A corresponds to the $N$ molecules that stacked onto the nanotubes during the functionalization and gave rise to the band B. Therefore, the changes in absorption of peak A and peak B are given by : $\Delta \alpha ^{A} = - N \epsilon^A ; \Delta \alpha ^{B} = N \epsilon^B$.
 We thus deduce :
 $$\frac{\epsilon^A}{\epsilon^B} = |\frac{\Delta \alpha ^A}{\Delta \alpha ^B}| \simeq 2 \pm 0.3 $$
 
 This factor 2 does not correspond to a true decrease of the absorption cross-section of the bound TPP but rather to a change in the orientational average. Actually, when the TPP molecules are bound to a nanotube they can absorb light only when the electric field is parallel to the tube axis, due to strong depolarization effects \cite{Roquelet2012}.
 Therefore, the orientation averaged cross-section of bound TPP is only 1/3 of the in-plane cross-section. In total, we obtain $\frac{\epsilon_{TPP}^{free}}{\epsilon_{TPP}^{bound}}=\frac{<\sigma_{TPP}^{free}>_{or}}{<\sigma_{TPP}^{bound}>_{or}}=2$ in agreement with the measurements. We thus deduce $\sigma_{TPP}^{//, free} = \sigma_{TPP}^{//, bound}$ : the in-plane absorption cross-section of TPP (that is relevant for the transfer ratio $R$) is not modified by the binding to the nanotube within a 15\% error bar.

\section*{global fitting procedure}

\begin{figure*}
 	\centering
 		\includegraphics[width=\textwidth]{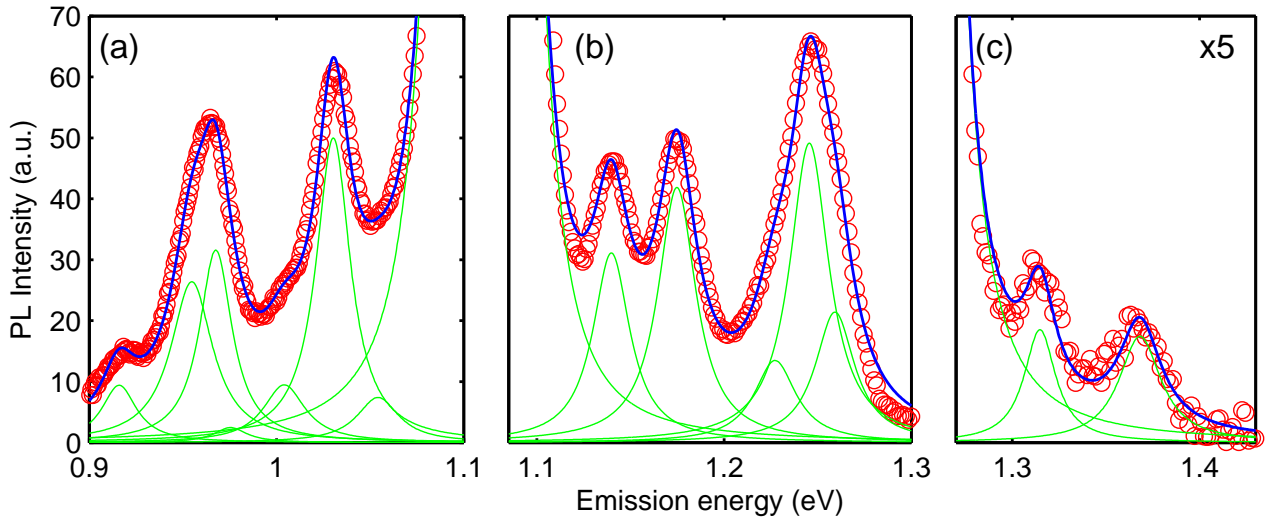}
 	\caption[width=1\textwidth]{(a) (b) and (c) Photoluminescence spectrum of the HiPCO SWNT/TPP compounds suspension with a resonant transfer excitation at 2.82 eV (red circles). Blue line : Global fit to the spectrum consisting of the sum of Lorentzian lines for each chiral species observed in the sample (green lines).}
 	\label{fig:fits}
\end{figure*}

To evaluate the $(n,m)$ PL intensities from the emission spectra, and deduce the values of $R$ discussed in the main text, a global fitting procedure is performed. Figure~\ref{fig:fits} shows an example of a fit along with the corresponding spectral data, for an excitation energy corresponding to the energy transfer resonance. 

To assess the dozen of species measurable with our experimental setup, the fit process is split into three spectral zones in order to obtain a better precision for dim species (Figures~\ref{fig:ppesde}a, \ref{fig:ppesde}b, \ref{fig:ppesde}c).
In a first step, the energy and width of each $(n,m)$ PL line are evaluated by means of a Lorentzian fit to the PL spectrum excited at the corresponding $S_{22}$ excitation energy, where this specific PL line is prominent. Values ranging from 22 to 28 meV for the full width at half maximum (FWHM) are found. From these individual fits, we construct a model function consisting of the sum of all the Lorentzian lines with free amplitude but fixed central energy and width. Finally, a fit to the PL spectra at each excitation energy of the map is performed using this model function, the only free parameters being the relative amplitudes of the peaks. This yields the intensities $I^{NT}_{Soret}$ and $I^{NT}_{S_{22}}$ for the evaluation of $R$ for each $(n,m)$ species. This global fit was performed on the PL maps of several samples of similarly functionalized compounds suspensions (5 from HiPCO material and 2 from CoMoCAT material), which gave reproducible results. The error bars of $R$ are evaluated considering 
both the fitting uncertainties and the statistical variations.     
 
We chose to fit the PL lines to Lorentzian profiles because such profiles gave the best agreement with experimental spectra. This is consistent with the fact that the PL linewidth of individual nanotubes at room temperature is of the order of 25~meV \cite{Tsyboulski2007} showing that inhomogeneous broadening (that would rather yield Gaussian profiles) is not prominent. 
That being said, we note that the values of $R$ evaluated for the spectrally well isolated PL lines, namely for the $(6,5)$, $(8,6)$, $(7,5)$ and $(10,2)$ species, is hardly sensitive to the fitting profile. This reduced set of data already gives the trend described in the main text, meaning that our results do not depend on the chosen fitting profiles.
The data obtained for the $(10,3)$, $(7,3)$ and $(9,2)$ minority species, though included in the fitting procedure of the HiPCO based spectra, are not considered in the discussion since they present much larger error bars. They are also relatively sensitive to the chosen fitting profile. This is due to their much weaker PL signal, on the order or even lower than the tails of the neighboring lines.
In addition, not all bright species could be used in this study because some emission energies are too close to each other (on the same column in the PL map), preventing us from singling out the intensity of the energy transfer resonance associated with each species. 
This restriction holds for the $(9,4)$, $(7,6)$ and $(8,4)$ species, which contributions add up in a single emission line at 1.09~eV. We note however that the intensity of the transfer resonance associated with these three chiral species is equal (within a 6\% uncertainty) to the sum of the three intrinsic intensities weighted with the $R$ values extrapolated from Figure~3 of the paper (for a global energy transfer resonance at 1.09~eV normalized to 1, we find $S_{22}$ resonances of resp. 0.28, 0.30 and 0.18. The extrapolated $R$ values are resp. 1.0, 1.3 and 1.5, leading to a weighted sum of 0.94). This restriction also holds for the $(12,1)$, $(10,5)$ and $(11,1)$ species. Their lower contributions are taken into account as small corrections in the evaluation of $R$ for respectively the $(8,6)$, $(9,5)$ and $(8,7)$ species (typically a ten percent reduction of $I^{NT}_{Soret}$).

Note that the PL due to the intrinsic absorption of SWNTs at 2.82~eV represents about 15\% of the total signal. Its value was deduce by applying the same global fitting procedure to non-functionalized nanotubes (Figure~\ref{fig:Map-2D}). This intrinsic excitation of SWNTs was subtracted from $I^{NT}_{Soret}$ for the calculation of $R$. At the same time, we took into account the contribution of the nearby $S_{33}$ transition for the (9,7) and (8,7) species.

\section*{diameter dependence of $\sigma_{S_{22}}$ and $\phi_{PL}$}

\begin{figure*}
 	\centering
 		\includegraphics[scale=0.5]{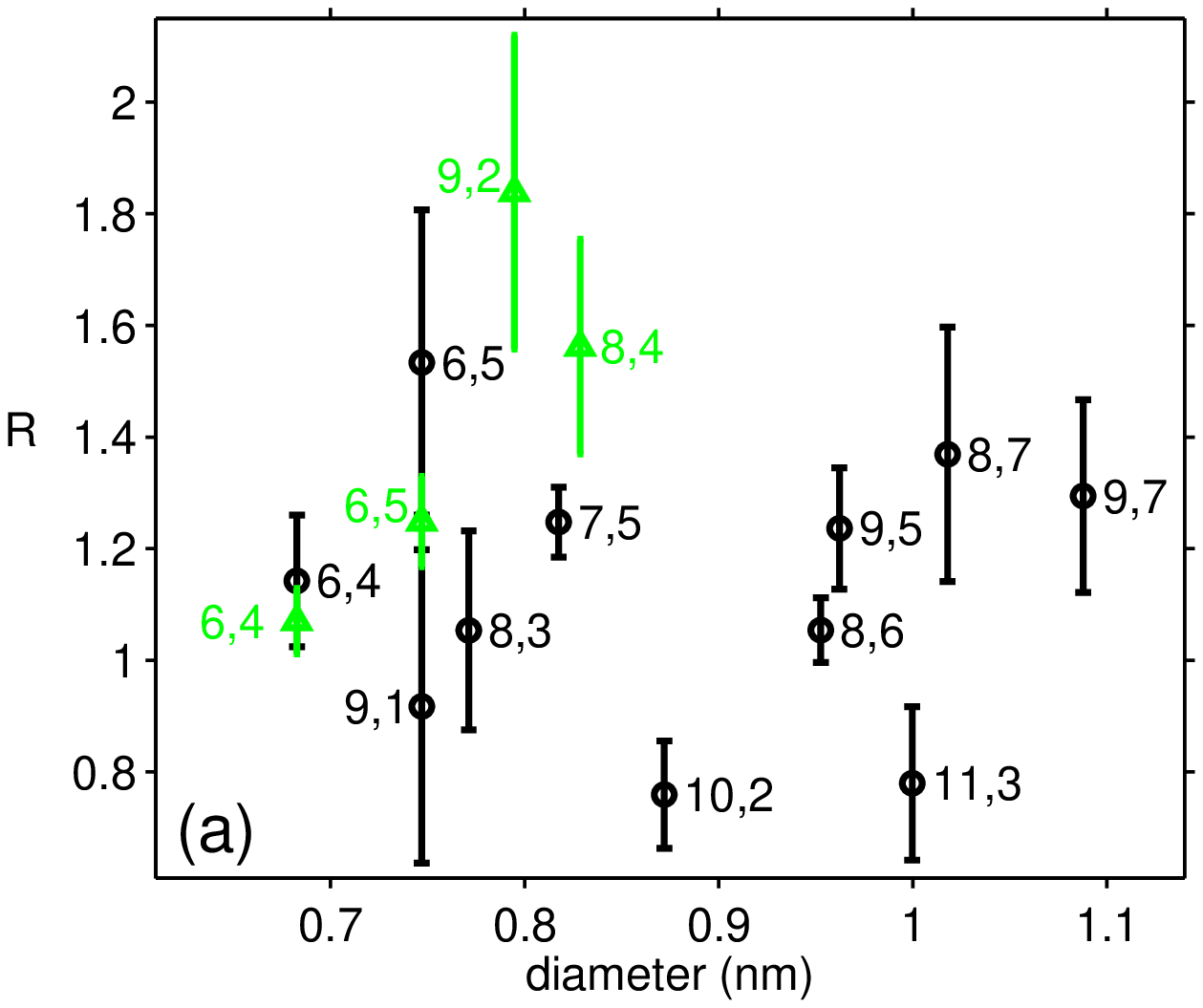}
 		\includegraphics[scale=0.5]{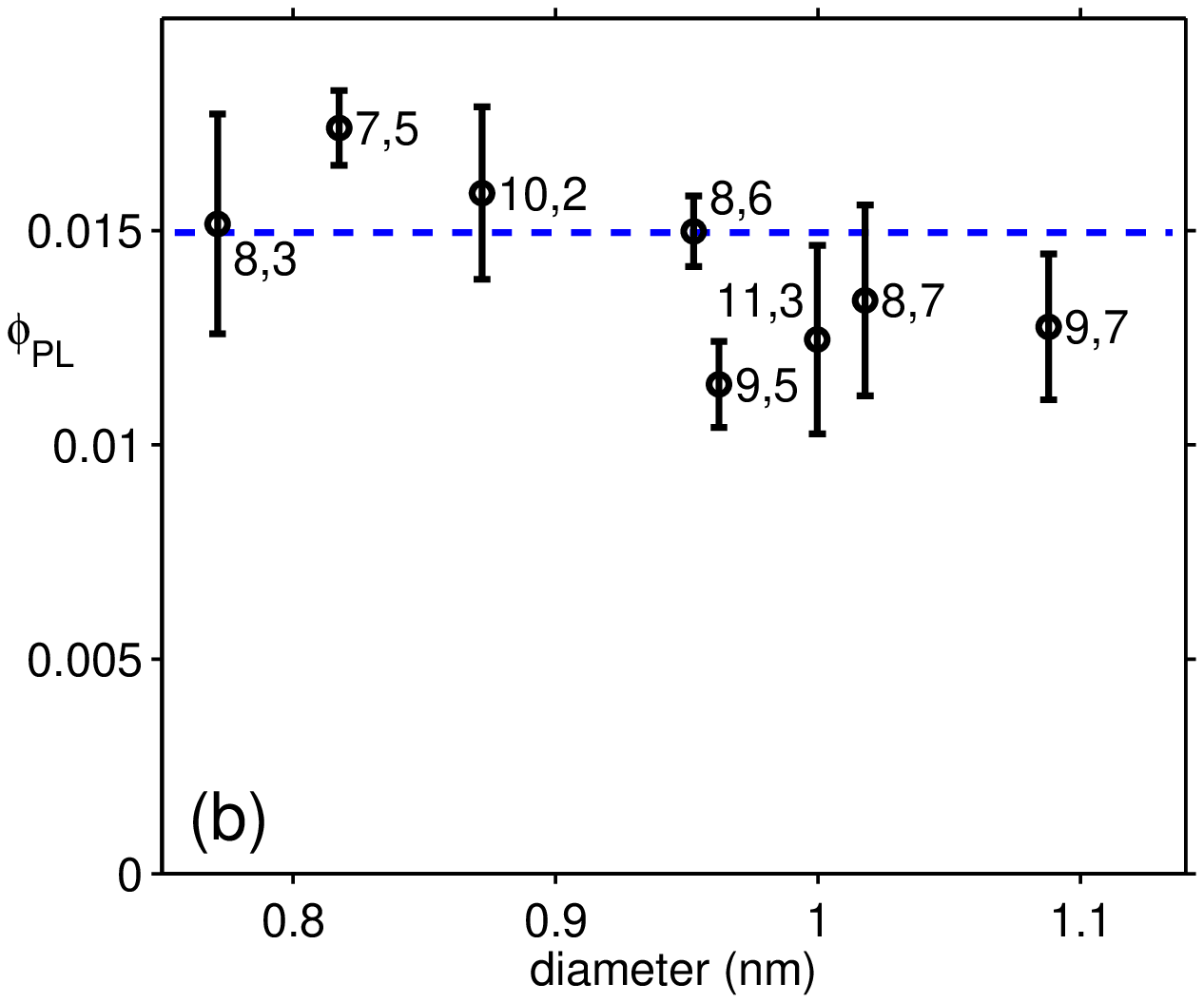}
 	\caption{(a) $R$ ratio as a function of the nanotube diameter evaluated for SWNT/TPP compound suspensions made from HiPCO (black circles) and CoMoCat (green triangles) sources. (b) Photoluminescence quantum yield $\phi_{PL}$ of the sample of ref.\cite{Tsyboulski2007}, evaluated using our estimate of $\sigma_{S_{22}}$, as a function of the nanotube diameter.}
 	\label{fig:R_phi_D}
 \end{figure*}

We observe no diameter dependence of $R$  and hence of $\sigma_{S_{22}}$ in our set of chiral species ($0.68 \ \mathrm{nm}< d_t< 1.1 \ \mathrm{nm}$, Figure~\ref{fig:R_phi_D}a). 

The PL quantum yield is deduced by dividing the action cross-section measured by Tsyboulski \textit{et al.} \cite{Tsyboulski2007} by our estimate of the absorption cross-section $\sigma_{S_{22}}$. No diameter (Figure~\ref{fig:R_phi_D}b) nor chiral angle dependence (see main text) of the PL quantum yield $\phi_{PL}$ is observed within the error bars. The average $\phi_{PL}$ is compatible with values reported in the literature on similar samples \cite{Carlson2007}.

All the experimentally measured data (absorption cross-sections $\sigma_{S_{22}}^{//}$ and PL quantum yield $\phi_{PL}$) are reported in the Table~\ref{table:phi_theta_diameter}.

 \begin{table*}
\setlength{\tabcolsep}{8pt}
\begin{center} 
\begin{tabular}{|c|c|c|c|c|c|}
 \hline
 $(n,m)$ & Type & $d_t$ (nm) & $\theta$ ($^\circ$) & $\sigma_{S_{22}}$ (nm$^2$/$\mu$m) & $\phi_{PL}$ (\%) \\ \hline \hline
$(6,4)$&I&0.68&23&410&\\ \hline
$(9,1)$&I&0.75&5&490&\\ \hline
$(8,3)$&I&0.77&15&430&1.5\\ \hline 
$(7,5)$&I&0.82&25&360&1.7\\ \hline
$(10,2)$&I&0.87&9&590&1.6\\ \hline
$(8,6)$&I&0.95&25&430&1.5\\ \hline
$(11,3)$&I&1.00&12&580&1.2\\ \hline
$(9,7)$&I&1.09&26&350&1.3\\ \hline \hline
$(6,5)$&II&0.75&27&330&\\ \hline
$(9,2)$&II&0.79&10&240&\\ \hline
$(8,4)$&II&0.83&19&290&\\ \hline
$(9,5)$&II&0.96&21&360&1.1\\ \hline
$(8,7)$&II&1.02&28&330&1.3\\ \hline

 \end{tabular}
\end{center}
\caption{Measured absorption cross-section at the $S_{22}$ transition ($\sigma_{S_{22}}$) and PL quantum yield ($\phi_{PL}$) deduced from Ref~\cite{Tsyboulski2007}, as a function of the chiral species.} 
\label{table:phi_theta_diameter}
\end{table*}

\section*{Species abundance evaluation}

The global fitting procedure described previously can be applied to the PL map of a regular suspension of non functionalized HiPCO nanotubes (Figure~\ref{fig:Map-2D}) to obtain the PL intensities associated with each chiral species. 

\begin{figure}[h!]
	\centering
		\includegraphics[width=0.50\textwidth]{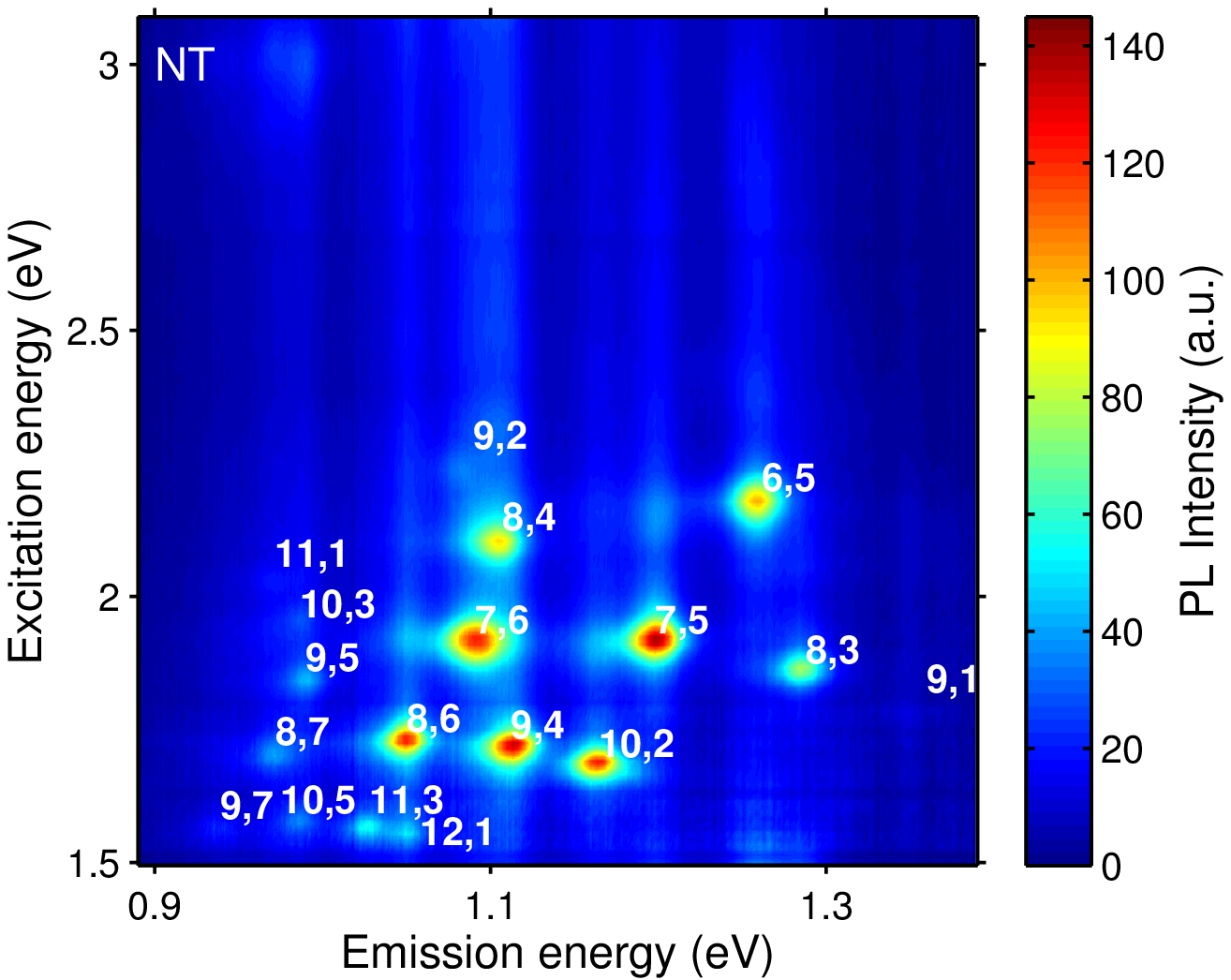}
	\caption{PL map of a HiPCO nanotube suspension. The spectra are normalized to a constant excitation photon flux.}
	\label{fig:Map-2D}
\end{figure}

\begin{figure*}[h!]
	\centering
		\includegraphics[width=\textwidth]{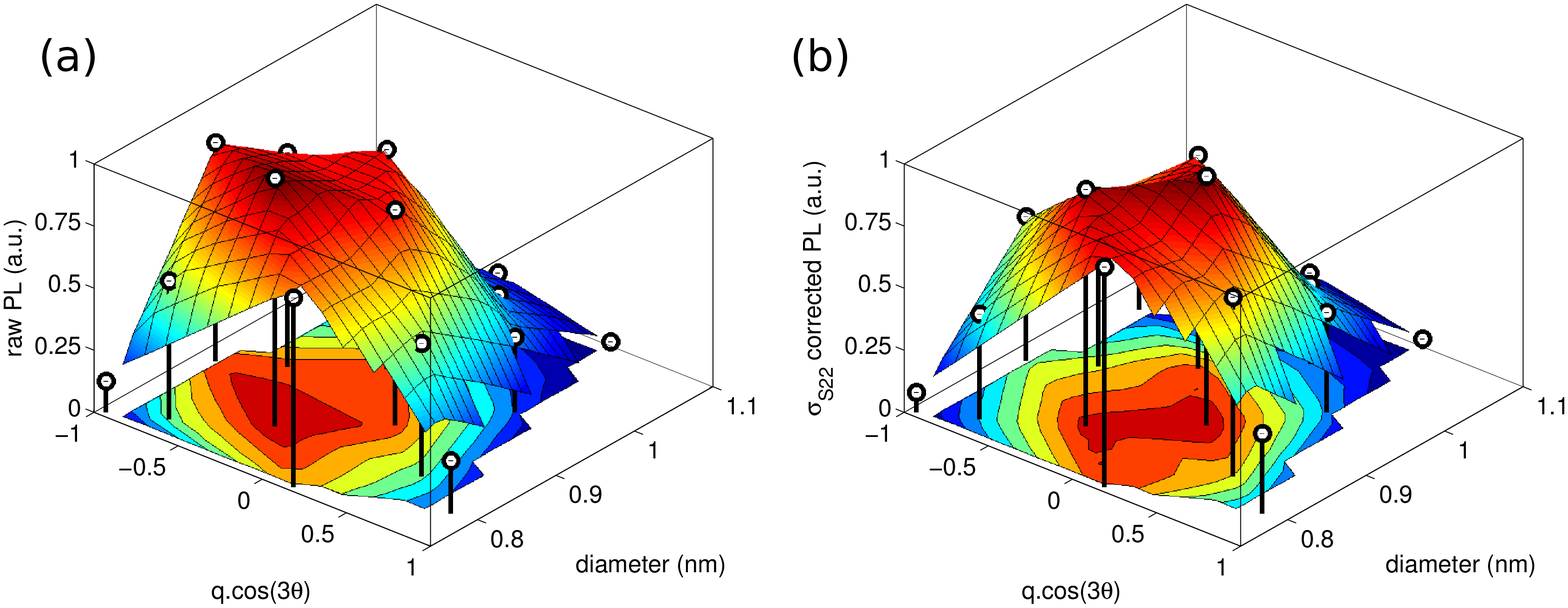}
	\caption{(a) Raw PL intensities extracted for 15 individual chiral species (black circles) from a HiPCO based suspension, as a function of the nanotubes geometrical parameters $q\cos(3\theta)$ and diameter. A linearly interpolated envelope is shown for the sake of clarity. (b) Similar data corrected from the chiral dependence of the absorption.}
	\label{fig:3D}
\end{figure*}

\begin{figure*}[h!]
	\centering
		\includegraphics[width=\textwidth]{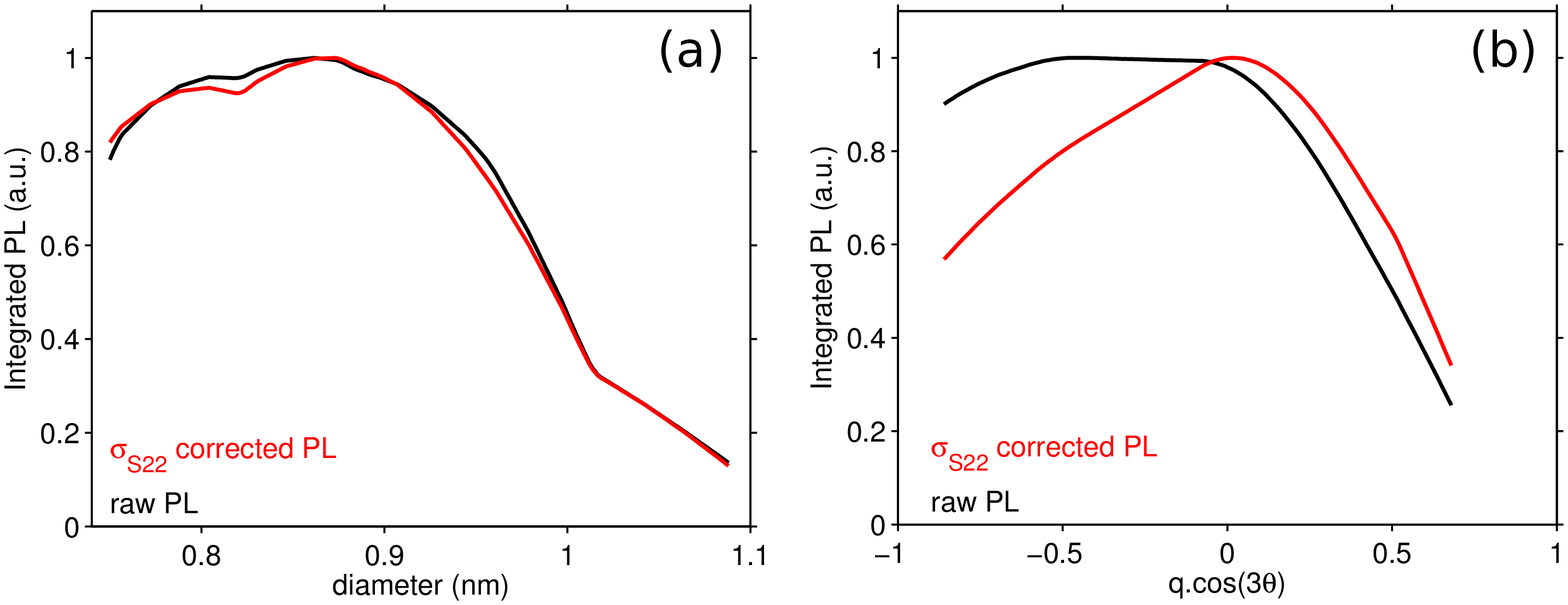}
	\caption{PL intensity distribution deduced from the enveloppe in Fig.~\ref{fig:3D}, as a function of (a) the diameter and (b) $q \cos 3 \theta$ : raw data (black line) and absorption corrected data (red line).}
	\label{fig:PLvsd+q}
\end{figure*}

This fitting procedure yields the PL intensities for a set of 15 individual species when excited at their $S_{22}$ resonance. These intensities are corrected from the spectral detection response of our setup. The relative variations of these raw PL intensities with respect to the geometrical parameter $q\cos(3\theta)$ and nanotube diameter are presented in Figure~\ref{fig:3D}a. A linearly interpolated envelope is displayed to help visualization. Similar studies have been reported in the literature to evaluate the relative abundance of the chiral species in a given sample. Both absorption and PL quantum yield were assumed not to depend on the chiral species \cite{Bachilo2002,Miyauchi2004}. 
With such an assumption, we find here a relatively symmetric diameter distribution, that reaches a maximum at about 0.85~nm and extends approximately from 0.75 to 0.95 nm (Figure~\ref{fig:PLvsd+q}a). The smaller diameter species are not well detected due to the limited spectral range measurable with our setup. On the contrary, the chiral angle distribution (Figure~\ref{fig:PLvsd+q}b) clearly indicates an asymmetric distribution with prominent type I ($q=-1$) nanotubes. One could conclude that the growth method used for this sample favors the type I near zigzag nanotubes and produces almost no type II near zigzag tubes. However, such a growth asymmetry would be very puzzling regarding the growth mechanism. 

Considering the chiral species dependence of the absorption evidenced in this study and the flat quantum yield, we can now trace back the chiral species abundance in a much more reliable way, by dividing the PL data by the absorption cross-section for each (n,m) species. The corrected distribution is shown in Figure~\ref{fig:3D}b and in Figures~\ref{fig:PLvsd+q}a, \ref{fig:PLvsd+q}b.

 Due to their higher absorption, the corrected PL intensities of type I near zigzag nanotubes are lowered while the ones for type II near zigzag tubes are increased owing to their lower absorption. This correction yields no significant change regarding the diameter distribution (Figure~\ref{fig:PLvsd+q}a) but has a major effect on the chiral angle distribution : it now yields an almost symmetric distribution with respect to the chiral family (Figure~\ref{fig:PLvsd+q}b). 
Note however that despite the correction factor the distribution still shows a clear tendency to a higher abundance of near armchair species. We even notice that no zigzag species could be observed in this study. This could be related to the theoretically predicted vanishing PL quantum yield for zigzag species \cite{Oyama2006}. This point is questionable though, since the model also predicts for instance a 5 fold reduced PL quantum yield for the (8,6) species as compared to the (8,3) whereas our data show no significant difference.

\bibliography{biblio-sigmaS22-SI}

\end{document}